\documentclass[12pt]{article}
\usepackage{amssymb}
\usepackage{amsmath}
\usepackage{graphicx,psfrag,epsf}
\usepackage{enumerate}
\usepackage{tabularx}
\usepackage{url} 
\usepackage{tikz}
\usepackage{bm}
\usepackage{float}
\usepackage{multirow}
\usepackage{threeparttable}
\usepackage{booktabs}
\usepackage{color}

\usepackage{array}
\usepackage{latexsym}
\usepackage{amsfonts}
\usepackage{psfrag}
\usepackage{authblk}

\usetikzlibrary{arrows,calc,positioning,decorations.pathreplacing}

\usepackage{natbib,color}

\newtheorem{assum}{Assumption}

\begin{document}

\def\spacingset#1{\renewcommand{\baselinestretch}%
{#1}\small\normalsize} \spacingset{1}


  \title{\bf Sampling from Networks: Respondent-Driven Sampling}
\author[1]{Mamadou Yauck\thanks{
  E-mail: \textit{mamadou.yauck@mcgill.ca}}}
\author[1]{Erica E. M. Moodie}
\author[1]{Herak Apelian}
\author[1]{Marc-Messier Peet}
\author[2]{Gilles Lambert}
\author[3]{Daniel Grace}
\author[4]{Nathan J. Lachowsky}
\author[5]{Trevor Hart}
\author[1]{Joseph Cox}
\affil[1]{McGill University, Montreal, Quebec, Canada}
\affil[2]{Institut national de santé publique du Québec, Montreal, Quebec, Canada}
\affil[3]{University of Toronto, Toronto, Ontario, Canada}
\affil[4]{ University of Victoria, Victoria, British Columbia, Canada}
\affil[5]{ Ryerson University, Toronto, Ontario, Canada}

  \maketitle



\bigskip
\begin{abstract}
Respondent-Driven Sampling (RDS) is a variant of link-tracing, a sampling technique for surveying hard-to-reach communities that takes advantage of community members' social networks to reach potential participants. While the RDS sampling mechanism and associated methods of adjusting for the sampling at the analysis stage are well-documented in the statistical sciences literature, methodological focus has largely been restricted to estimation of population means and proportions, while giving little to no consideration to the estimation of population network parameters. As a network-based sampling method, RDS is faced with the fundamental problem of sampling from population networks where features such as homophily (the tendency for individuals with similar traits to share social ties) and differential activity (the ratio of the average number of connections by attribute) are sensitive to the choice of a sampling method. Though not clearly described in the RDS literature, many simple methods exist to generate simulated RDS data, with specific levels of network features, where the focus is on estimating simple estimands. However, the accuracy of these methods in their abilities to consistently recover those targeted network features remains unclear. This is also motivated by recent findings that some population network parameters (e.g.~homophily) cannot be consistently estimated from the RDS data alone \citep{Crawford17}.

In this paper, we conduct a simulation study to assess the accuracy of existing RDS simulation methods, in terms of their abilities to generate RDS samples with the desired levels of two network parameters: homophily and differential activity. The results show that (1) homophily cannot be consistently estimated from simulated RDS samples and (2) differential activity estimates are more precise when groups, defined by traits, are equally active and equally represented in the population. We use this approach to mimic features of the Engage Study, an RDS sample of gay, bisexual and other men who have sex with men in Montréal.

\end{abstract}
\noindent%
{\it Keywords: Network data; Respondent-driven sampling; Simulation.}
\vfill
\newpage
\spacingset{1.45} 

\section{Introduction}
Hard-to-reach communities such as sex workers, people who use drugs, or men who have sex with men may be unwilling to participate in a research study, often because of social stigma \citep{Hec97}.
Members of such communities are, however, often connected through a social network. Current sampling strategies such as snowball sampling \citep{Good61}, a variant of link-tracing sampling, take advantage of those social relationships to reach members of the study population who are not easily accessible to researchers. The snowball sampling mechanism is non-probabilistic, and can result in selection or sampling biases that may affect the accuracy of any estimates calculated via the produced samples \citep{Gile11, Gile15b}. Because of this problem, snowball samples are often referred to as `convenience samples' that lack any valid basis for inferential methods whose results might generalize to the underlying population of interest \citep{Bier81}.

\textit{Respondent-Driven Sampling} (RDS) was introduced \citep{Hec97} as a form of link-tracing sampling that aimed to combine the advantages of probabilistic sampling and snowball sampling, with the idea that `those best able to access members of hidden populations are their own peers'. In RDS, the study recruitment protocol leads to the generation of a large number of recruitment \textit{waves}, with each successive participant being asked to recruit additional participants starting from initial ‘seed’ participants that are purposively selected. RDS offers several advantages over traditional link-tracing methods. First, the RDS recruitment occurs through a number of waves, allowing the process to sample further from the seeds and reducing the dependence of the final sample on the initial sample. Second, allowing respondents to recruit their peers reduces the confidentiality concerns associated with listing respondents' social network contacts. Finally, RDS adjusts all analyses for the (self-reported) social connectivity of participants, thereby ensuring the resulting estimates account for the relative over- (or under-) sampling of those members of the community who are more (or less) socially connected and thus more (or less) likely to be invited to participate in the study.

Current research on RDS is focused mainly on estimating population means and proportions \citep{Gile15b, Gile18} while overlooking the estimation of population network parameters. In fact, being a network-based sampling method, RDS is faced with the fundamental problem of sampling from population networks where features such as homophily and differential activity, two measures of social `connectedness' of individuals with similar traits are sensitive to the choice of a sampling method \citep{Cost03}. Moreover, sensitivity analyses in current RDS research is concerned with population estimators (means and proportions) with regard to network and sampling assumptions \citep{Gile15b}, for various levels of network features such as homophily and differential activity, while failing to address the accuracy of RDS simulation methods in terms of their abilities to recover those specific network features. This is a crucial problem since \cite{Crawford17} showed that homophily cannot be consistently estimated given the observed RDS data alone, or without additional assumptions on the structure of the observed RDS network. Moreover, \cite{shalizi2013consistency} showed that applying Exponential Graph Random Models (ERGM) \citep{harris2014introduction}, a class of generative network models that are routinely used in RDS studies to simulate population networks, to a partially observed network raises consistency issues for a large class of network parameters. 

In this paper, we define homophily and differential activity, our target parameters, as functions of the population network and covariate values, and assess the accuracy of their estimates from simulated RDS samples. In the next section, we introduce RDS as a network-based sampling technique and describe its sampling mechanism. In section \ref{sec:methodo}, we present the current methodology for simulating RDS samples from population networks and, in Section \ref{sec:simulation}, present a simulation study to assess the accuracy and precision of homophily and differential activity estimates. In Section \ref{sec: casestudy}, we mimic features of a real-world dataset about HIV transmission among gay, bisexual and other men who have sex with men (GBM), recruited in Montréal using RDS.

\section{Background}
Simulation of RDS samples requires two steps. First, a population network with known characteristics and relational structures must be simulated. Second, an RDS sample with prespecified characteristics is drawn from the population network \citep{Spil18, Gile10}. In this section, we describe the required assumptions and key characteristics of the population and the sampling process.

\subsection{The population network and the inferential targets}\label{sec:popnetwork}
Suppose a target population of $N$ individuals. Following \cite{Hec97}, \cite{Sal04} and \cite{Gile11}, we assume that the individuals in the population, or \textit{nodes} in the network, are connected by social ties.
\begin{assum}
The social network connecting members of the target population exists and is an undirected graph $G = (V, E)$ with no parallel edges or self-loops.
\end{assum}
\textit{Parallel edges} refer to two edges with the same end vertices. An edge is called \textit{self loop} if it has the same vertex as both its end vertices.The elements in the set $V$ of vertices are the $|V|=N$ individuals (or nodes) in the population, while the edges in $E$ represent social ties between members of the population. Let $\bm{y}$ be an $N\times N$ adjacency matrix representing the dual relationships in the network, with elements $y_{ij}=y_{ji}$ indicating the presence of an edge between nodes $i$ and $j$.
 Each node $i \in V$ is assigned a degree $d_i=\sum_{j=1}^N y_{ij}$, defined as the number of edges connected to that node. We assume that the degrees over the entire network are distributed according to $D=\left(D_0,\ldots,D_K\right)$, where $D_k=\sum_{i=1}^N I\left(d_i=k\right)$ represents the number of nodes with degree $k$, $K$ represents the maximum degree, and $D$ can be viewed as a population-level frequency table. The degree distribution is subjected to the following constraint of consistency: $\sum_{k=0}^K kD_k=\sum_{i=1}^N d_i=2|E|$, where $|E|$ is the number of edges in G. Finally, we define $\bar{d}=\frac{1}{N}\sum_{i=1}^N d_i$ as the mean degree of the network.

\begin{figure}[H]
\begin{center}
\includegraphics[scale=.85]{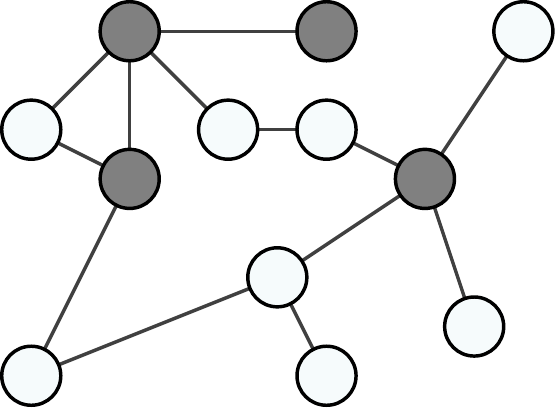}
\end{center}
%
\caption{A population network with minority group in grey, and with $N=12$, $p=0.33$, $\bar{d}=2.16$, $D_a=1.16$, $R=0.40\,(h=-0.20)$.}
\label{fig:popnet}
\end{figure}

Let $\bm{z}$ be a $1\times N$ vector of two-valued attributes $z_i \in \{0,\;1\}$ for the $i-th$ individual in the population, and $p=\frac{1}{N}\sum_{i=1}^N z_i$  the proportion of individuals with attribute $z_i=1$. We focus on two-valued nodal attributes, but the results presented in this paper can be extended to categorical attributes.
We define two inferential targets (and features of the network structure on nodal attributes): \textit{differential activity} ($D_a$) and \textit{homophily} ($h$). Differential activity is equal to the ratio of mean degrees by attribute,
\begin{equation*}
D_a=\frac{\sum_{i=1}^N z_id_i}{\sum_{i=1}^N (1-z_i)d_i} \frac{1-p}{p},
\end{equation*}
and can be thought of as a measure of the relative `social connectedness' of those with and without the trait. Homophily is the tendency for individuals with similar traits to share social ties.
Several measures of homophily have been used in the RDS literature \citep{Gile18}. In this work, two related measures of homophily in the population network will be considered. Let $z_{ji}=1$ if node $i$ has attribute $j$. A fairly intuitive measure of homophily is Newman's assortativity coefficient for categorical attributes \citep{Newm02}, defined as
\begin{equation}\label{assortnewman}
h=\frac{\sum_{j\in\{0,\;1\}}p_{jj}-\sum_{j\in\{0,\;1\}}p_{j\bullet}p_{\bullet j}}{1-\sum_{j\in\{0,\;1\}}p_{j\bullet}p_{\bullet j}},
\end{equation}
where
$$
p_{jk}=\frac{1}{|E|}\sum_{i=1}^N \sum_{l=1}^N y_{il}z_{ji}\left(1-z_{kl}\right),\,j\neq k,\,i \neq l
$$
is the proportion of network edges linking a node of attribute $j$ to one of attribute $k$, 
\begin{equation*}
p_{00}= \frac{1}{|E|}\sum_{i=1}^N \sum_{l<i} y_{il}\left(1-z_{ji}\right)\left(1-z_{kl}\right),\mbox{~~and~~}p_{11}=  \frac{1}{|E|}\sum_{i=1}^N \sum_{l<i} y_{il}z_{ji}z_{kl}.
\end{equation*}
Further, $p_{j\bullet}=\sum_{k\in \{0,\,1\}} p_{jk}$ is the fraction of network edges originating from a node of attribute $j$ and $p_{\bullet j}$ is the proportion of  edges terminating in a node of category $j$. Newman's assortativity coefficient ranges from $-1$ to $1$. A value of $h=1$ indicates a perfect assortative mixing, $h=0$ depicts no assortative mixing, while at $h=-1$ the network is said to be dissortative.
The level of homophily can also be measured by 
$R=p_{11}/p_{10}$ ($p_{10}\neq 0$) \citep{Gile11}. These two metrics are related since, for the undirected graph $G$,  Equation (\ref{assortnewman}) can be expressed as
\begin{equation}
h=\frac{R}{1+R}-\frac{2}{1+\eta\left(1+2R\right)},
\end{equation}
where $\eta=\frac{1}{D_a}\frac{1-p}{p}$.
Even though formulation (\ref{assortnewman}) is easier to interpret, we will primarily rely on the measure $R$ in the simulations of Section \ref{sec:simulation} for ease of implementation. An illustration of a network with a single nodal attribute and prespecified levels of prevalence, homophily and differential activity is presented in Figure \ref{fig:popnet}.

%
%
%

\subsection{The RDS process}\label{sec:RDSsample}
RDS is a network-based sampling technique in which members of a hidden community reach across their personal social network to recruit other members into a study. We begin by describing how RDS works in practice, before considering how this translates to the implementation of RDS in simulation. The RDS process can be described as follows:
\begin{enumerate}\setlength{\itemsep}{-6pt}
\item Sampling starts with the selection of a fixed number of nodes, or \textit{seeds}, in the network. In practice, seeds are often chosen purposefully so as to be as heterogeneous as possible with respect to nodal attributes. Seeds recruit members to the study by $(i)$ inviting them to participate, and $(ii)$ giving their invited social contacts a \textit{coupon} that they return to the researcher, so that the researcher can track the social links in the recruitment process. Coupons have unique identifying numbers and/or letters to link recruiters to their recruits.  Seeds, and successive participants, all receive a fixed number of coupons. 
\item Each seed recruits further participants, up to the total number of coupons received. Individuals may only participate once in the study.
\item Each successive implementation of Step 2 is called a \textit{wave}. The recruitment continues, through a number of waves, until the desired sample size is reached. 
\end{enumerate}

To consider how the above practical implementation of study recruitment can be formalized, we first state the assumptions on the graphical structure of RDS.

\begin{figure}[H]

\begin{center}
\includegraphics[scale=.85]{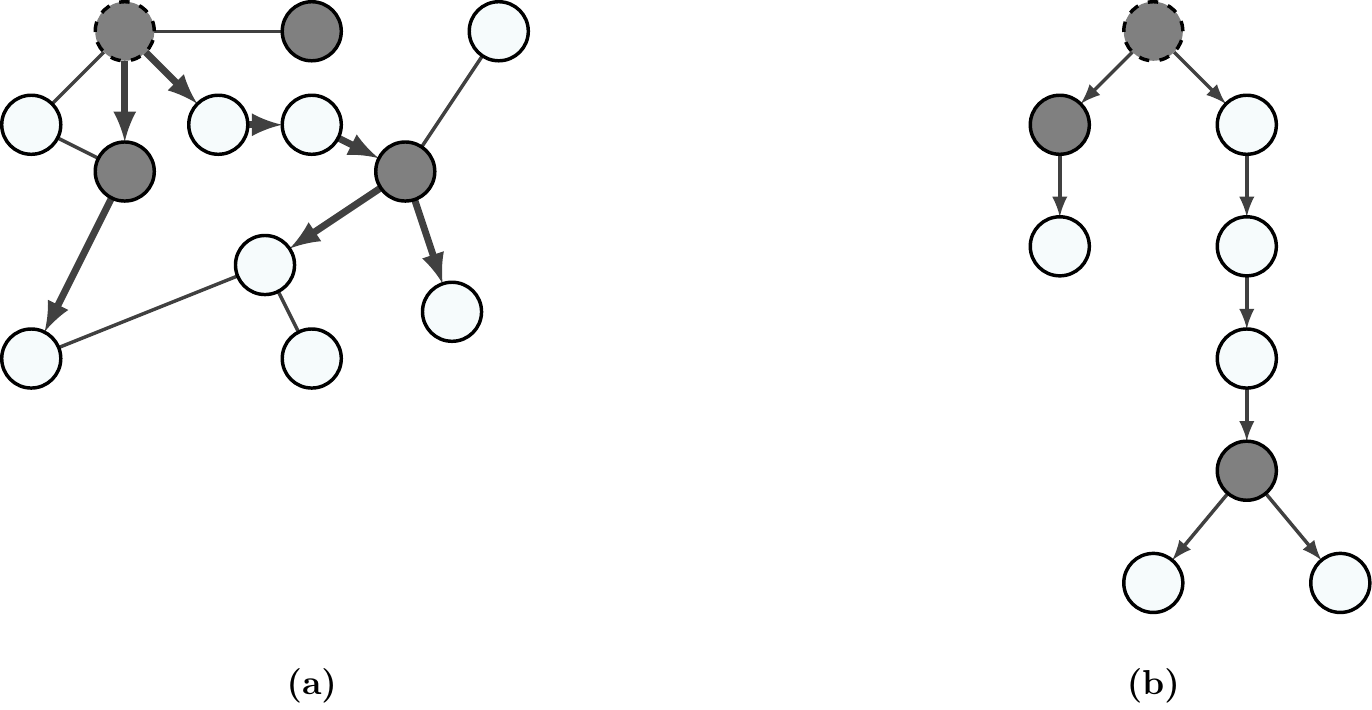}
\end{center}

%
%
%
%

\caption{Illustration of the RDS process  with one seed (dashed circle), two coupons per sampled node and a sample of size $8$. The minority group is represented in grey. The RDS recruitment chain navigating through the population network is illustrated in \textbf{(a)} and the resulting RDS recruitment graph is illustrated in \textbf{(b)}, indicating 5 waves of recruitment.}
\label{fig:RDSsampling}
\end{figure}

\begin{assum}
The RDS recruitment is conducted across edges of the undirected graph G.
\end{assum}
\begin{assum}
No node in the social network can be sampled more than once.
\end{assum}
Successive samples are obtained by sampling among the remaining unsampled neighbours (social contacts) of sampled nodes in the population. Each sampled node selects up to a predetermined fixed number of unsampled neighbors. The recruitment process stops when the desired sample size is reached. An example of the RDS process is illustrated in Figure \ref{fig:RDSsampling}.

The general procedure to simulate an RDS sample is as follows:
\begin{enumerate}\setlength{\itemsep}{-6pt}
\item Sample, without replacement, a fixed number of seeds from the $N$ nodes of the population network. The selection of seeds is either dependent or independent of $\bm{z}$. When the selection is conditional on nodal attributes, there is potential `seed bias' induced by the selection of the initial sample, especially when there is strong homophily on $\bm{z}$. 
\item For each seed, sample up to a number of nodes bounded by the number of coupons, without replacement, from among their unsampled neighbors. If the selection depends on $\bm{z}$, then there is \textit{differential recruitment} in the sampling process.
\item Repeat step 2 until the desired sample size is reached. In practice, the target sample size is usually linked to the diversity of the RDS samples with respect to the population characteristics upon which sampling focuses.
\end{enumerate}
One key advantage of the RDS process over other link-tracing sampling methods is that through a fixed number of waves, the dependence of the final sample on the initial sample is reduced or eliminated. 
 From a practical standpoint, one needs to decide on the number of seeds to sample and coupons to distribute, as the latter is inversely proportional to the number of waves of sampling \citep{WHORDS2013} under the assumption that recruits accept an invitation randomly. When homophily on $\bm{z}$ is weak, the process will likely reach equilibrium after a small number of waves \citep{Gile10}. In this case, the decision on whether to distribute a small (large) number of coupons to a large (small) number of seeds will not have a great impact on the final sample. When the network is highly clustered, sampling should be conducted in a way that allows a broad range of individuals to recruit from their networks through many waves. This can be achieved by distributing a small number of coupons to an initial sample as diverse as possible with respect to the characteristics of the target population.

\section{Methodology}\label{sec:methodo}
In this section, we describe a classical approach to simulating a population network given specific features, and to generating RDS samples from the simulated network.

\subsection{Simulating a population network}\label{sec:SimNetwork}

A common approach to generating the social network of the population is by simulating using Exponential Random Graph models \citep{harris2014introduction}, a class of generative models based on exponential family distribution theory for modeling network dependence. Let $\bm{Y}$ be the random adjacency matrix for the population network. The joint distribution of its elements is defined as
\begin{equation}\label{ergm}
\mbox{P}\left(\bm{Y}=\bm{y}|\bm{z}\right)=\frac{\mbox{exp}\left\lbrace \bm{\theta} g\left(\bm{y},\bm{z}\right) \right\rbrace }{\bm{\kappa\left(\bm{\theta}\right)}},
\end{equation}
where $g\left(\bm{y},\bm{z}\right) $ is a vector of network statistics and $\bm{\theta}$ its corresponding vector of parameters, $\bm{\kappa\left(\bm{\theta}\right)}=\sum_{\bm{y}} \mbox{exp}\left\lbrace \bm{\theta} g\left(\bm{y},\bm{z}\right) \right\rbrace$ is a normalizing constant. The main structural features of the network are fully captured in model (\ref{ergm}) by choosing statistics to represent density, degree distribution by attribute and homophily on nodal attributes. The sufficient statistic for the network density is $g_0(\bm{y})=\sum_{i=1}^N\sum_{j=1}^Ny_{ij}=|E|$. The statistics for the degree distribution by attribute are obtained by counting the number of times a node with such attributes appears in an edge:
\begin{equation*}
g_1\left(\bm{y},\bm{z}\right) = \sum_{j=1}^N \sum_{k<j} y_{jk}z_jz_k+\sum_{j=1}^N \sum_{k=1}^N y_{jk} z_j\left(1-z_k\right),\mbox{~~and~~}
\end{equation*}

\begin{equation*}
g_2\left(\bm{y},\bm{z}\right) =\sum_{j=1}^N \sum_{k<j} y_{jk}\left(1-z_j\right)\left(1-z_k\right)+ \sum_{j=1}^N \sum_{k=1}^N y_{jk}z_j\left(1-z_k\right).
\end{equation*}
The sufficient statistics for homophily are represented by the joint distribution of the node and neighbour's attribute (also called \textit{mixing matrix}). For a $2\times2$ mixing matrix of an undirected graph, one needs to specify the following statistics:
\begin{equation*}
g_3\left(\bm{y},\bm{z}\right) = \sum_{j=1}^N \sum_{k<j} y_{jk}z_jz_k \mbox{~~and~~} g_4\left(\bm{y},\bm{z}\right)= \sum_{j=1}^N \sum_{k<j} y_{jk}\left(1-z_j\right)\left(1-z_k\right).
\end{equation*}
By expressing (\ref{ergm}) in terms of the conditional log-odds of a tie between two nodes, one can show that $\bm{\theta}$ represents the log-odd of a tie conditional on all others \citep{harris2014introduction}.
 If we assume that $Y_{ij}$ and $Y_{kl}$ are independent for any $\left(i,\,j\right) \neq \left(k,\,l\right)$, then $g\left(\bm{y},\bm{z}\right) =g\left(\bm{y}\right)$=$\sum_{i=1}^N\sum_{j=1}^Ny_{ij}$ and (\ref{ergm}) reduces to
\begin{equation}\label{ergm-bernou}
\mbox{P}\left(\bm{Y}=\bm{y}|\bm{z}\right)=\frac{\mbox{exp}\left( \theta |E| \right) }{\bm{\kappa\left(\theta\right)}},
\end{equation}
which corresponds to the simplest random graph model, also called the Bernouilli model \citep{durrett_2006}. In model (\ref{ergm-bernou}) the probability of a tie between any two nodes, $\frac{\exp\left(\theta\right)}{ 1+\exp\left(\theta\right) }$,  is the density of the network.

ERGMs can be fitted via $\textsf{statnet}$ \citep{handcock:statnet}, a suite of $\textsf{R}$ packages, including $\textsf{ergm}$ \citep{Hunter08}, $\textsf{sna}$ \citep{Carter08-1} and $\textsf{network}$ \citep{Carter08-2}, for the modeling of network data. The structural features of the population network are included as \textit{terms} in the function $\textsf{ergm}$ of the same package.
Homophily on nodal attributes and differential activity are specified in the function call to $\textsf{ergm}$ by using terms $\textsf{nodematch}$ and $\textsf{nodecov}$ respectively.

If there is more than one nodal covariate, a two-step procedure is used to control both the relationship between covariates and the network structure on each covariate. First, we generate covariates with known dependence and marginal distributions using the package $\textsf{GenOrd}$ \citep{Ale17}. Then compute network statistics corresponding to homophily and differential activity for each covariate as inputs for $\textsf{ergm}$. 

Once the model is fully specified and fitted, one can simulate an undirected network from the distribution of all undirected networks that are consistent with the target statistics.

\subsection{Simulating a study that applies RDS to the population network}\label{sec:SimNetwork}
RDS samples are drawn from the synthetic undirected population networks using a two-step procedure. First, $s$ nodes are sampled (sequentially) without replacement as \textit{seeds}. In this work, we assume that there is no `seed bias' as the selection regime of the initial sample does not depend on $\bm{z}$. Successive sampling \textit{waves} are obtained by sampling sequentially, and without replacement, up to $c$ nodes from among the unsampled neighbours of each selected node. We assume that their is no differential recruitment. The process is halted once the sample size reaches $n$. The parameters of the RDS sampling process are defined in Table \ref{table:parameter_RDS}.

\begin{table}[H]
\caption{Parameters of the RDS sampling process.}
\begin{center}
  \setlength\extrarowheight{-2pt}
\begin{tabular}{lcc} \hline
{Parameter} & {} & {Notation}   \\ \hline
& \textbf{Population network} \\
Population size &        &  $N$   \\
Prevalence &        & $p$  \\
Mean degree&        &  $\bar{d}$   \\
Differential activity &        &  $D_a$  \\
Homophily &        &  $R$\\ \\
 & \textbf{RDS sample} \\
Number of seeds &        &   $s$  \\
Number of coupons&        &  $c$   \\
Sample size&        &  $n$ \\
\hline
\end{tabular}
\end{center}
\label{table:parameter_RDS}
\end{table}

\section{Simulations}\label{sec:simulation}
The goal of the simulation study is to assess the accuracy and precision of homophily and differential activity estimates when RDS samples drawn from population networks using the methodology described in Section \ref{sec:methodo}. 
\subsection{Simulation setup}
 We simulated population networks, with a single nodal attribute $\bm{z}$, from which RDS samples were drawn
 for the set of characteristics defined in Table \ref{table:parameter_values}.
\begin{table}[H]
\caption{Parameters of simulated networks and RDS samples.}
\begin{center}
  \setlength\extrarowheight{-2pt}
\begin{tabular}{lcc} \hline
{Parameter} & {} & {Values}   \\ \hline
& \textbf{Population network} \\
$N$  &        &  $1000$  \\
 $p$ &        & $0.1,\,0.5,\,0.8$  \\
 $\bar{d}$ &        & $99.9$  \\
 $D_a$ &        & $0.5,\,1,\,4$   \\
 $R$ &        & $1,\,5$\\ \\
 & \textbf{RDS sample} \\
  $s$ &        & $5$  \\
 $c$ &        & $2$   \\
 $n$ &        & $200,\,400,\,800$\\
\hline
\end{tabular}
\end{center}
\label{table:parameter_values}
\end{table}
We performed 500 simulation runs for each set of characteristics and computed crude estimates of homophily ($h$) and differential activity ($D_a$) from each simulated RDS sample. For each estimator $\hat \theta$ ($\hat h$ or $\hat D_a$) of the target parameter $\theta$ ($h$ or $D_a$), we computed the relative bias as
\begin{equation*}
RB(\hat \theta)=\frac{\hat\theta-\theta}{\theta}.
\end{equation*}
The results are presented in the next section.

\subsection{Differential activity and homophily biases}
\begin{figure}[H]
\centering
\includegraphics[scale=.9]{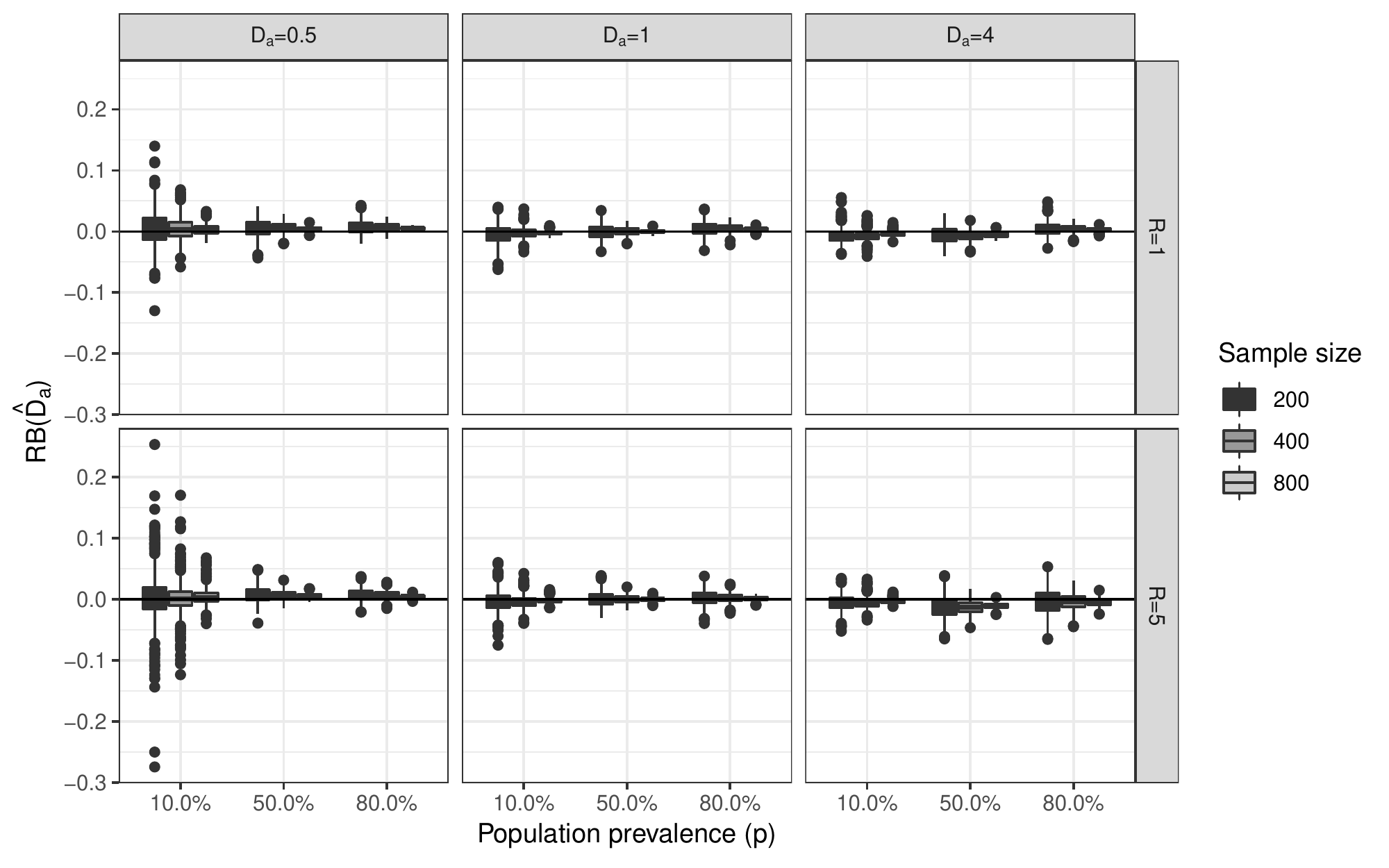}
\caption{ Differential activity bias for a population network of size $N=1000$ and mean degree $\bar{d}=99.9$. The x-axis shows the levels of population prevalence, $p=0.1,\,0.5,\,0.8$. The y-axis visualizes the relative bias of the differential activity estimator $\hat{D}_a$. The $2\times 3$ grid depicts the levels of differential activity ($D_a=0.5,\,1,\,4$) and homophily ($R=1,\,5$). The boxplots for each set of characteristics are presented for three sample sizes ($n=200,\,400,\,800$). The number of seeds and coupons are set to $s=5$ and $c=2$, respectively.}
\label{fig:da.bias}
\end{figure}

\begin{figure}[H]
\centering
\includegraphics[scale=.9]{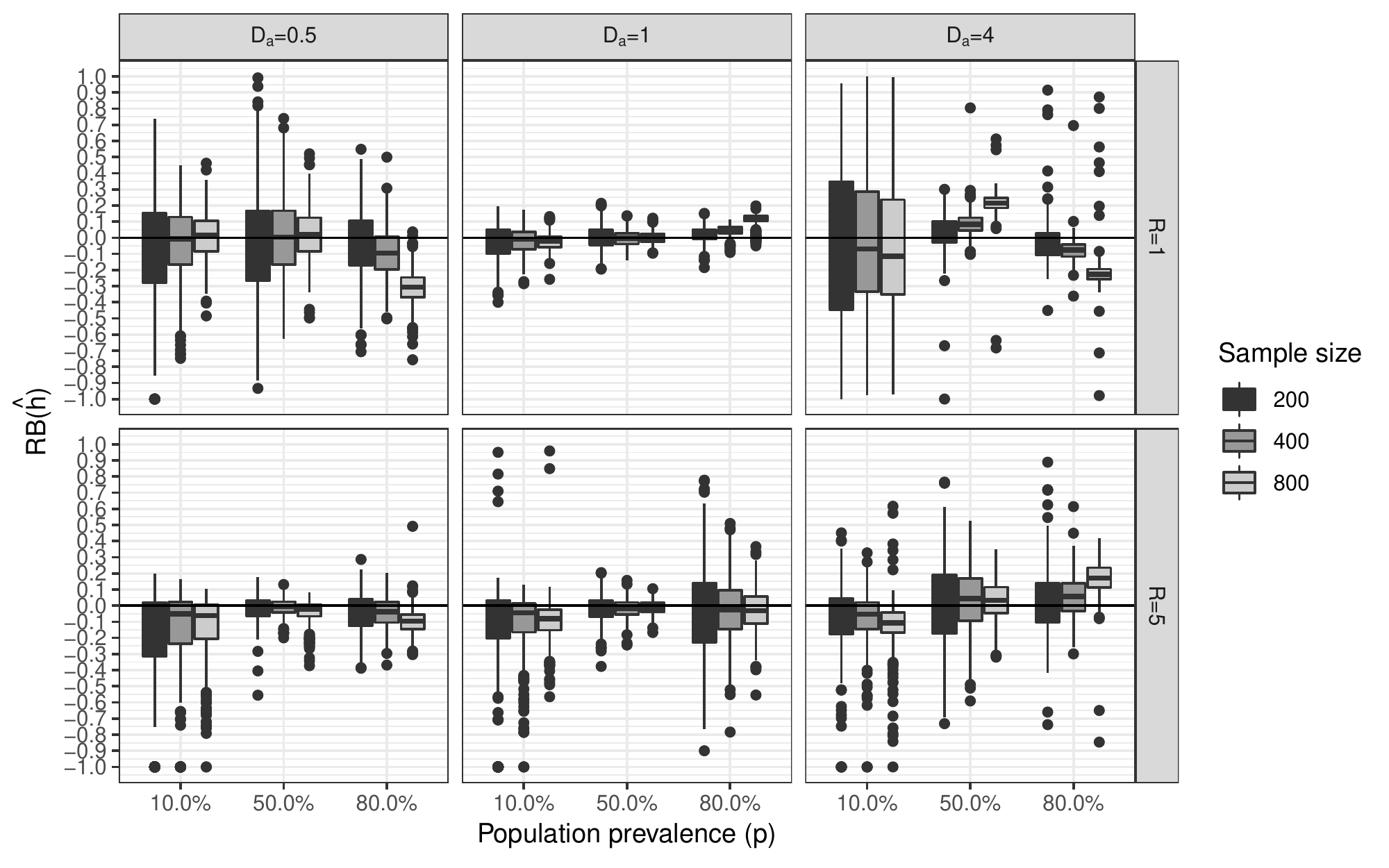}
\caption{ Homophily bias for a population network of size $N=1000$ and mean degree $\bar{d}=99.9$. The x-axis shows the levels of population prevalence, $p=0.1,\,0.5,\,0.8$. The y-axis visualizes the relative bias of the homophily estimator $\hat{h}$. The $2\times 3$ grid depicts the levels of differential activity ($D_a=0.5,\,1,\,4$) and homophily ($R=1,\,5$). The boxplots for each set of characteristics are presented for three sample sizes ($n=200,\,400,\,800$). The number of seeds and coupons are set to $s=5$ and $c=2$, respectively.}
\label{fig:hmp.bias}
\end{figure}

The distribution of the relative biases  for each set of network and sample characteristics are illustrated in Figures \ref{fig:da.bias}-\ref{fig:hmp.bias} for the estimators of  differential activity and homophily, respectively. All cases are compared to the setting in which $R=1$, $D_a=1$ and $p=50\%$. In this setting, there is negligible bias in all estimators regardless of sample size.

In Figure \ref{fig:da.bias}, we show the distribution of the relative bias for the level of activity in the network. The bias is negligible in all cases. There is more variability when the $\bm{z}$-absent minority group is twice as active as the $\bm{z}$-present group, which decreases as more nodes are sampled.


Figure \ref{fig:hmp.bias} shows the bias distribution for homophily. When both groups are equally active, the bias is negligible when $R=1$. The bias is higher on average when the $\bm{z}$-present group, whether in the minority or the majority, is four times more active than the $\bm{z}$-absent group. Furthermore, the estimation of homophily exhibits more variability as the difference in the level of activity between the $\bm{z}$-present and the $\bm{z}$-absent groups increases.  An important observation is that the estimator deteriorates as the sampling fraction increases.  This result was demonstrated by \cite{lin2013sampling} while comparing different sampling techniques on social networks. This can be explained by the fact that the RDS recruitment tree is only a partially observed network, with missing ties between nodes within the sample. Further, network characteristics that are (primarily) functions of the type of connections such as homophily cannot be consistently estimated given the observed RDS data alone (\citealt{Crawford17,shalizi2013consistency}). 


Overall, network parameter estimates from RDS samples are more accurate when both groups ($\bm{z}$-present and $\bm{z}$-absent) are equally active and equally represented in the population. For differential activity, the estimation becomes more accurate as the sample fraction increases. The RDS  process performs well in terms of recovering the true level of homophily for small to medium sampling fractions but deteriorates as the sampling fraction increases, confirming previous findings that it is hard to estimate homophily consistently when the true network is only partially observed through the RDS recruitment tree.

\section{Case Study}\label{sec: casestudy}
\begin{figure}[H]
\begin{center}
\includegraphics[scale=.7]{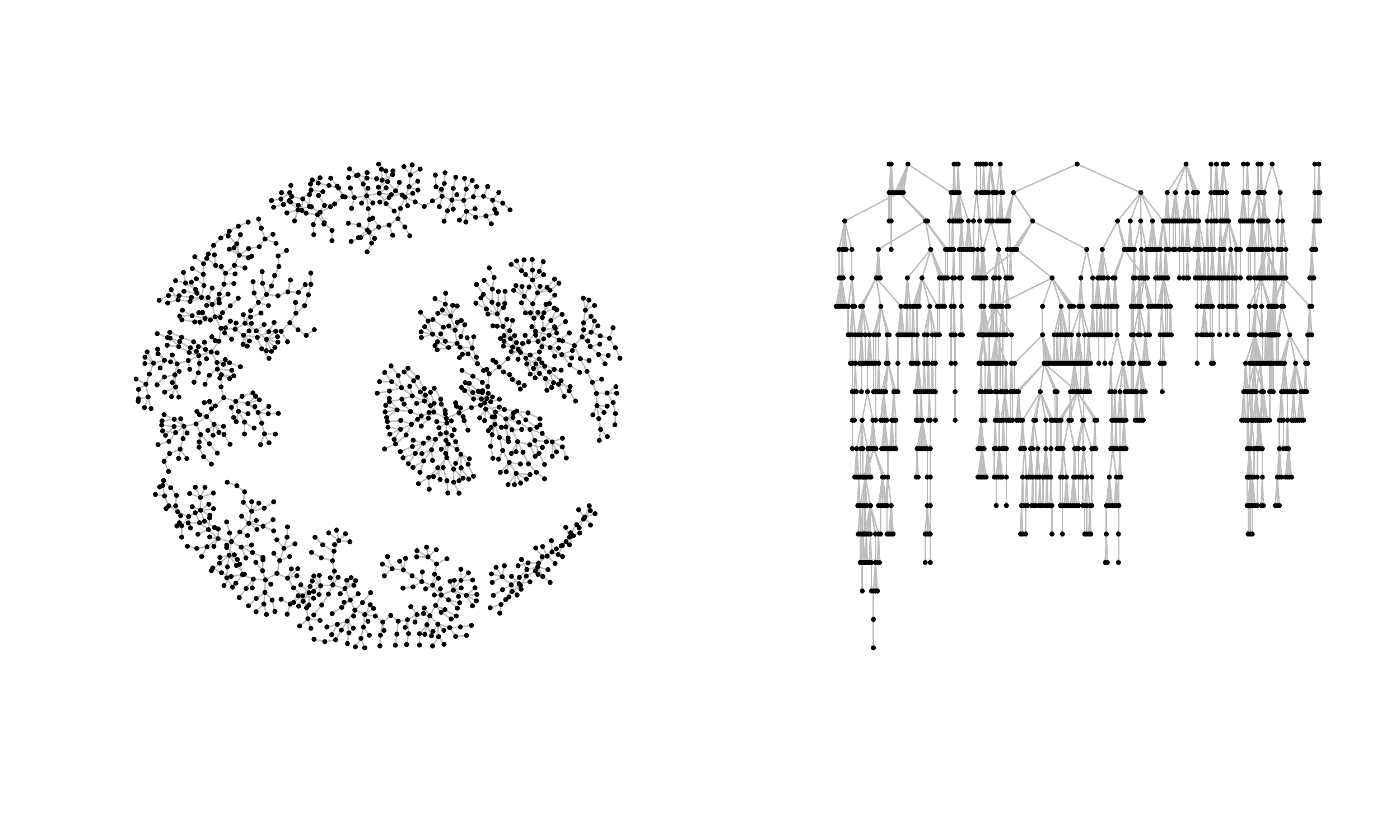}
\end{center}
\vspace*{-3cm}
\caption{ Representation of the observed RDS network among $n=1179$ gay, bisexual and other men who have sex with men (GBM) in Montréal, 2018. On the left, a sphere representation of the same network. On the right, a representation of the recruitment tree in which nodes are aligned by wave.}
\label{fig:engagenet}
\end{figure}

We now turn to data collected through the Engage study, a national cross-sectional study undertaken in three large Canadian cities, Montréal, Toronto and Vancouver. The main goal of the Engage study is to determine the individual, social and community-level factors that impact HIV and STI transmission and related behaviours  among GBM \citep{DRSP19}; for the Engage Montréal recruitment network, see Figure \ref{fig:engagenet}. In this example, we focus on data collected in Montréal, and aim to generate, as a proof of concept, synthetic samples that mimic a subset of the observed data in terms of key covariate features including differential activity and homophily.

Participants in the Engage study were recruited using RDS. The process started with the selection of $s=27$ seeds with ages ranging from 16 to 80, who mostly identified as French Canadian (17), English Canadian (1),  European (4), Caribbean (1), Arab (1), South-East Asian (1) and mixed (2), with four participants living with HIV. Seeds were selected following a formative assessment and community mapping, and to be as heterogeneous as possible with respect to the diversity (e.g., HIV status, ethnicity) of the GBM community. At the end of their interview, participants received $c=6$ uniquely identified coupons, along with monetary and non-monetary incentive (complete STI screening), to recruit their peers into the study population. The study was  conducted from February 2017 through June 2018 for a total of $n=1179$ GBM recruits. Approximately $55\%$ of the recruited individuals who were given coupons, including $6$ seeds, did not recruit anyone, while $82\%$ of the effective recruiters recruited between $1$ to $3$ members.

Descriptive statistics of the RDS sample are displayed in Table \ref{table:descstat}. About $33\%$ of respondents were aged less than 30 years, seven out of ten were born in Canada, two-thirds were French or English Canadian, $30\%$ had a high school diploma or lower, and around $58\%$ earned less than $30\,000\$$ in annual income. Around $86\%$ of respondents described themselves as gay and two out of five reported being in a relationship with a main partner. In the past six months, almost $14\%$ of GBM recruits declare using crack cocaine, and less than $1\%$ reported using a syringe used by someone else in the past six months. About two-thirds of respondents reported having anal sex without a condom during the past six months, and almost $17\%$ reported that they were living with HIV.

The mean degree of the observed RDS network is $\bar{d}=51.77$. The level of homophily on covariates described in Table \ref{table:descstat} ranged from 0.08 (Use of a syringe used by someone else) to 0.46 (Age), depicting a small to moderate homophilic network on average, with respect to nodal attributes. The differential activity level ranged from 0.67 (Age) to 1.40 (Place of birth). Respondents with a college degree and those with a lower diploma were almost equally active (as measured in terms of their degree, or number of social links), while respondents who reported living with HIV were $32\%$ more active than those who reported an HIV negative status.

The goal of this example is to simulate RDS samples with network and sample characteristics similar to those of the (simulated) Engage population network. First, we simulate $1000$ population networks with three nodal covariates: condomless anal sex (CAS), currently in a relationship with a main partner (CIR) and HIV status (HIV+)) using the method described in Section \ref{sec:methodo}. The true size of the population is set to $N=40400$ \citep{ISQ16}.

\begin{figure}[H]
\centering
\includegraphics[scale=.8]{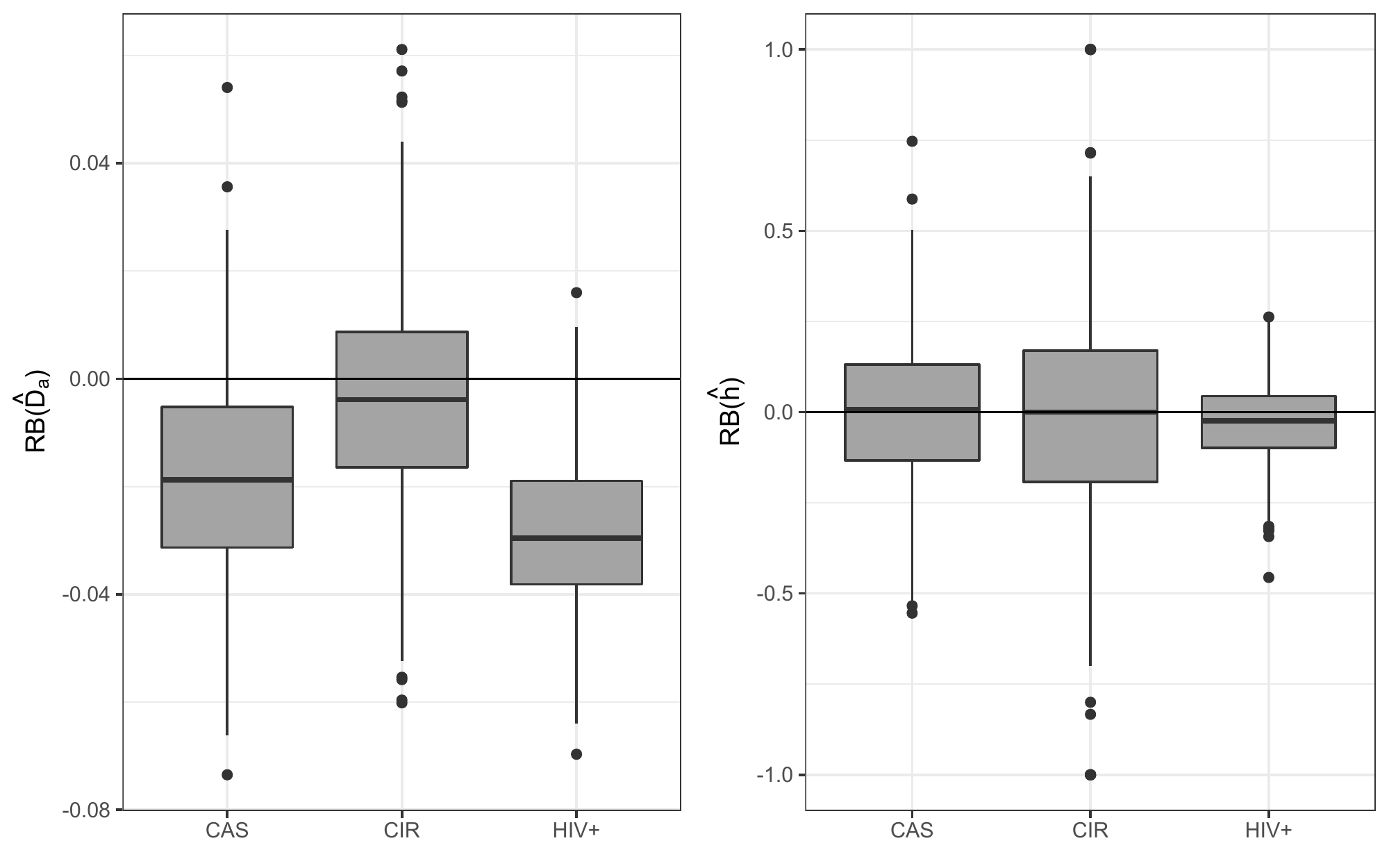}
\caption{ Differential activity ($D_a$) and homophily ($h$) biases for the underlying population network of gay, bisexual and other men who have sex with men (GBM) recruits. The nodal attributes are \textit{Condomless anal sex} (CAS), \textit{Currently in a relationship with a main partner} (CIR) and \textit{HIV positive} (HIV+).}
\label{fig:engage.appli}
\end{figure}

For each nodal attribute, we estimate the prevalence by adjusting for each individual's reported social network size using the RDS-II estimator \citep{Hec02}, the levels of homophily and differential activity, and take these values to be the true values in the population. Then, for each simulated population network, we simulate an RDS sample with characteristics that mimic those of the empirical RDS sample. The summary statistics of the network and RDS sample characteristics are presented in Table \ref{table:parameter_Engage} (see Appendix). The association matrix for the three nodal attributes is presented in Table \ref{table:correlation} and displayed in the Appendix. There is a positive and significant association between having sex without a condom and being in a relationship with a main partner. Having a positive HIV status is not significantly associated with having sex without a condom or being in a relationship. We compute relative biases for differential activity and homophily on each of the three nodal covariates. Figure \ref{fig:engage.appli} shows the bias distribution for differential activity and homophily.

The relative bias for the level of activity is small to negligible for the three covariates. The bias is more pronounced for HIV+ ($-2.9\%$ on average) as this group is $32\%$ more active than the HIV- group. The bias is negligible for the nodal attribute CIR ($<1\%$ on average) as both groups are approximately equally represented ($p=44\%$) and equally active ($D_a=0.95$).

The homophily bias is small (for HIV+) to negligible (for CAS and CIR)  on average. Overall, this result was expected as groups for both CAS and CIR are approximately neutral, equally active and equally represented in the network. Although small ($-2.56\%$ on average), the magnitude of the relative homophily bias for HIV status was also expected as the HIV+ group is $32\%$ more active, exhibits medium homophilic behavior ($h=0.38$) and represents $17\%$ of the network population.


\begin{table}[H]
\caption{Descriptive statistics of the RDS sample of $n$=1179 gay, bisexual and other men who have sex with men (GBM) recruits in Montréal: number, percent, homophily and differential activity.}
\vspace*{-0.4cm}
\begin{center}
 \setlength\extrarowheight{-1pt}
\footnotesize
\begin{tabular}{lllllcccc} \hline
 & & &&& {\bf $n$} & {\bf \%}& {\bf $h$}& {\bf $D_a$}   \\
\textbf{Socio-demographic characteristics}& & &&&&& &\\
\,\,\,\,\,\,Age less than 30 &   &&& &  $384$   &  $32.6$& $0.46$ & $0.67$\\
\,\,\,\,\,\,Born in Canada&     & &&& $820$    & $69.6$&$0.35$  & $1.40$\\
\,\,\,\,\,\,Not French or English Canadian&   &&& &  $445$   &  $37.8$&  $0.32$&$0.71$\\
\,\,\,\,\,\,Highest diploma lower than college's&   &&& &  $352$   &  $29.9$& $0.23$ &$0.98$\\
\,\,\,\,\,\,Less than $30\,000\$$ in annual income&   &&& &  $678$   &  $57.5$ & $0.20$&$0.77$\\ \\
\textbf{Sexuality}& & &&&&& &\\
\,\,\,\,\,\,Describe oneself as gay &   &&& &$1016$   &  $86.2$ &$0.27$ &$1.35$\\
\,\,\,\,\,\,Currently in a relationship with a main partner &   &&& &$508$   &$43.1$   & $0.09$&$0.95$\\
\,\,\,\,\,\,Anal sex without a condom during the past 6 months &   &&& &$761$   &$64.5$  &$0.17$ &$1.18$\\ \\
\textbf{Drug Use}& & &&&&& &\\
\,\,\,\,\,\,Use of crack cocaine&   &&& & $158$  &$13.4$   &$0.27$ &$1.27$\\
\,\,\,\,\,\,Use of a syringe used by someone else &   &&& &$68$   &$5.77$   &$0.08$ &$1.16$\\ \\
\textbf{Health Status}& & &&&& & &\\
\,\,\,\,\,\,HIV positive &   &&& &  $200$ &$16.9$   &$0.38$ &$1.32$\\
\hline
\end{tabular}
\end{center}
\label{table:descstat}
\end{table}

\section{Discussion}
	The simulation of RDS samples from population networks is an important methodological issue. While current research is focused on estimating population means (e.g. ~prevalence), there has been little to no study on the accuracy of simulation methods for RDS in their abilities to recover important network features such as homophily, differential activity and mean degree. This paper showed, via a simulation study, that $(i)$ network features such as homophily cannot be consistently recovered from simulated RDS samples, as highlighted in previous theoretical findings (\citealt{Crawford17,shalizi2013consistency}), and $(ii)$ estimates of differential activity are less precise when there is a difference in the level of activity between two groups of attributes. 
	
	 This is an important first step in assessing simulation methods that are currently used in RDS studies. This result will be particularly useful to RDS methodologists, who aim to provide new inferential tools or validate the approaches currently being used in practice.

\appendix
\section*{Appendix}

\begin{table}[H]
\caption{Characteristics of the Engage network population and RDS sample.}
\begin{center}
  \setlength\extrarowheight{-2pt}
\footnotesize
\begin{tabular}{lcccc} \hline
{Parameter} & {} & {Estimated}  & {} &{$95\%$ CI} \\ 
{} & {} & {Value}  & {} &{} \\ \hline
& \textbf{Population network} \\
Population size &        &  $40400$ &&   \\
Mean degree&        &  $16.63$  &  & \\
Prevalence ($\%$)&        &  &  &\\
\,\,\,\,\,\,\,\textit{Condomless anal sex in the past six months}& &$57.9$  & & $[52.7,\,63.0]$\\
\,\,\,\,\,\,\,\textit{Currently in a relationship} & &$43.9$   & &$[38.8, \,49.0]$ \\
\,\,\,\,\,\,\,\textit{HIV positive}& &$12.7$  & &$[9.3, \,16.0]$\\\\
 & \textbf{RDS sample} \\
Number of seeds &        &   $27$    && - \\
Number of recruits&        &     \\
\,\,\,\,\,\,\,\,\,\,\,0& &$651$   && -\\
\,\,\,\,\,\,\,\,\,\,\,1 & &$236$  && -\\
\,\,\,\,\,\,\,\,\,\,\,2& &$117$   &&-\\
\,\,\,\,\,\,\,\,\,\,\,3& &$81$    &&-\\
\,\,\,\,\,\,\,\,\,\,\,4 & &$49$  &&-\\
\,\,\,\,\,\,\,\,\,\,\,5& &$27$  &&-\\
\,\,\,\,\,\,\,\,\,\,\,6& &$18$  &&-\\
Sample size&        &  $1179$  &&-\\
\hline
\end{tabular}
\end{center}
\label{table:parameter_Engage}
\end{table}

\begin{table}[H]
\caption{(Pearson) correlation matrix of three nodal covariates for the Engage RDS sample. Unweighted and weighted correlations are displayed, with weighted correlations in parenthesis.}
\begin{center}
 \begin{threeparttable}
  \setlength\extrarowheight{0pt}
\footnotesize
\begin{tabular}{l|ccc}
{} & {1. CAS } & {2. CIR}  & {3. HIV+ } \\ \hline
1. Condomless anal sex (CAS)&   1     & $0.104$\tnote{***} \,\,\,\,\, ($0.115$\tnote{***}\,\,\,\,\,\,\,)& $0.023$ ($0.018$)\\
2. Currently in a relationship (CIR)&        &  1  & $0.046$ ($0.002$) \\
3. HIV positive (HIV+)&        &  & 1 \\
\hline
\end{tabular}
 \begin{tablenotes}
            \item[***] p-value$<0.001$.
        \end{tablenotes}
 \end{threeparttable}
\end{center}
\label{table:correlation}
\end{table}

\subsection*{Acknowledgment}
	The authors would like to thank the Engage study participants, office staff, and community engagement committee members, as well as our community partner agencies REZO, ACCM and Maison Plein Coeur. The authors also wish to acknowledge the support of David M. Moore, Nathan J. Lachowsky and Jody Jollimore and their contributions to the work presented here. 
Engage/Momentum II is funded by the Canadian Institutes for Health Research (CIHR, TE2-138299), the CIHR Canadian HIV/AIDS Trails Network (CTN300), the Canadian Foundation for AIDS Research (CANFAR, Engage), the Ontario HIV Treatment Network (OHTN, 1051), the Public Health Agency of Canada (Ref: 4500370314), Canadian Blood Services (MSM2017LP-OD), and the Ministère de la Santé et des Services sociaux (MSSS) du Québec.

	 Erica E. M. Moodie acknowledges a chercheur boursier senior career award from the Fonds de recherche du Québec – Santé and a Discovery Grant from the Natural Sciences and Engineering Research Council (NSERC) of Canada (RGPIN-2019-04230).

\bibliographystyle{jasaauthyear}
{\small \bibliography{All-References} }

\begin{thebibliography}{25}
\newcommand{\enquote}[1]{``#1''}
\expandafter\ifx\csname natexlab\endcsname\relax\def\natexlab#1{#1}\fi
\expandafter\ifx\csname url\endcsname\relax
  \def\url#1{\texttt{#1}}\fi
\expandafter\ifx\csname urlprefix\endcsname\relax\def\urlprefix{URL }\fi
\providecommand{\bibinfo}[2]{#2}
\providecommand{\noopsort}[1]{}
\providecommand{\switchargs}[2]{#2#1}

\bibitem[{Barbiero and Ferrari(2017)}]{Ale17}
\bibinfo{author}{Barbiero, A.} and \bibinfo{author}{Ferrari, P.~A.}
  (\textbf{\bibinfo{year}{2017}}). \enquote{\bibinfo{title}{An r package for
  the simulation of correlated discrete variables}},
  \bibinfo{journal}{Communications in Statistics - Simulation and Computation}
  \textbf{\bibinfo{volume}{46}}, \bibinfo{pages}{5123--5140}.

\bibitem[{Biernacki and Waldorf(1981)}]{Bier81}
\bibinfo{author}{Biernacki, P.} and \bibinfo{author}{Waldorf, D.}
  (\textbf{\bibinfo{year}{1981}}). \enquote{\bibinfo{title}{Snowball sampling:
  problem and techniques of chain referral sampling.}},
  \bibinfo{journal}{Sociological Methods and Research}
  \textbf{\bibinfo{volume}{10}}, \bibinfo{pages}{141– 163}.

\bibitem[{Butts(2008{\natexlab{a}})}]{Carter08-2}
\bibinfo{author}{Butts, C.} (\textbf{\bibinfo{year}{2008}}{\natexlab{a}}).
  \enquote{\bibinfo{title}{network: A package for managing relational data in
  r}}, \bibinfo{journal}{Journal of Statistical Software, Articles}
  \textbf{\bibinfo{volume}{24}}, \bibinfo{pages}{1--36}.

\bibitem[{Butts(2008{\natexlab{b}})}]{Carter08-1}
\bibinfo{author}{Butts, C.} (\textbf{\bibinfo{year}{2008}}{\natexlab{b}}).
  \enquote{\bibinfo{title}{Social network analysis with sna}},
  \bibinfo{journal}{Journal of Statistical Software, Articles}
  \textbf{\bibinfo{volume}{24}}, \bibinfo{pages}{1--51}.

\bibitem[{Camirand \emph{et~al.}(2016)Camirand, Traoré, and Baulne}]{ISQ16}
\bibinfo{author}{Camirand, H.}, \bibinfo{author}{Traoré, I.}, and
  \bibinfo{author}{Baulne, J.} (\textbf{\bibinfo{year}{2016}}).
  \emph{\bibinfo{title}{L’Enquête québécoise sur la santé de la
  population, 2014-2015: pour en savoir plus sur la santé des Québécois}}.

\bibitem[{Costenbader and Valente(2003)}]{Cost03}
\bibinfo{author}{Costenbader, E.} and \bibinfo{author}{Valente, T.~W.}
  (\textbf{\bibinfo{year}{2003}}). \enquote{\bibinfo{title}{The stability of
  centrality measures when networks are sampled.}}, \bibinfo{journal}{Social
  Networks} \textbf{\bibinfo{volume}{25}}, \bibinfo{pages}{283– 307}.

\bibitem[{Crawford \emph{et~al.}(2017)Crawford, Aronow, Zeng, and
  Li}]{Crawford17}
\bibinfo{author}{Crawford, F.~W.}, \bibinfo{author}{Aronow, P.~M.},
  \bibinfo{author}{Zeng, L.}, and \bibinfo{author}{Li, J.}
  (\textbf{\bibinfo{year}{2017}}). \enquote{\bibinfo{title}{Identification of
  homophily and preferential recruitment in respondent-driven sampling}},
  \bibinfo{journal}{American Journal of Epidemiology}
  \textbf{\bibinfo{volume}{187}}, \bibinfo{pages}{153--160}.

\bibitem[{Durrett(2006)}]{durrett_2006}
\bibinfo{author}{Durrett, R.} (\textbf{\bibinfo{year}{2006}}).
  \emph{\bibinfo{title}{Erdös–Rényi Random Graphs}},
  \bibinfo{pages}{27–69}, Cambridge Series in Statistical and Probabilistic
  Mathematics (\bibinfo{publisher}{Cambridge University Press}).

\bibitem[{Gile(2011)}]{Gile11}
\bibinfo{author}{Gile, K.~J.} (\textbf{\bibinfo{year}{2011}}).
  \enquote{\bibinfo{title}{Improved inference for respondent-driven sampling
  data with application to hiv prevalence estimation.}},
  \bibinfo{journal}{Journal of the American Statistical Association}
  \textbf{\bibinfo{volume}{106}}, \bibinfo{pages}{135– 146}.

\bibitem[{Gile \emph{et~al.}(2018)Gile, Beaudry, Handcock, and Ott}]{Gile18}
\bibinfo{author}{Gile, K.~J.}, \bibinfo{author}{Beaudry, I.~S.},
  \bibinfo{author}{Handcock, M.~S.}, and \bibinfo{author}{Ott, M.~Q.}
  (\textbf{\bibinfo{year}{2018}}). \enquote{\bibinfo{title}{Methods for
  inference from respondent-driven sampling data.}}, \bibinfo{journal}{Annual
  Review of Statistics and Its Application} \textbf{\bibinfo{volume}{5}},
  \bibinfo{pages}{65– 93}.

\bibitem[{Gile and Handcock(2010)}]{Gile10}
\bibinfo{author}{Gile, K.~J.} and \bibinfo{author}{Handcock, M.~S.}
  (\textbf{\bibinfo{year}{2010}}). \enquote{\bibinfo{title}{Respondent-driven
  sampling: an assessment of current methodology.}},
  \bibinfo{journal}{Sociological Methodology} \textbf{\bibinfo{volume}{40}},
  \bibinfo{pages}{285--327}.

\bibitem[{Gile \emph{et~al.}(2015)Gile, Johnston, and Salganik}]{Gile15b}
\bibinfo{author}{Gile, K.~J.}, \bibinfo{author}{Johnston, L.~G.}, and
  \bibinfo{author}{Salganik, M.~J.} (\textbf{\bibinfo{year}{2015}}).
  \enquote{\bibinfo{title}{Diagnostics for respondent-driven sampling.}},
  \bibinfo{journal}{Journal of the Royal Statistical Society}
  \textbf{\bibinfo{volume}{178}}, \bibinfo{pages}{241– 269}.

\bibitem[{Goodman(1961)}]{Good61}
\bibinfo{author}{Goodman, L.~A.} (\textbf{\bibinfo{year}{1961}}).
  \enquote{\bibinfo{title}{Snowball sampling}}, \bibinfo{journal}{Annals of
  Mathematical Statistics} \textbf{\bibinfo{volume}{32}},
  \bibinfo{pages}{148--170}.

\bibitem[{Handcock \emph{et~al.}(2003)Handcock, Hunter, Butts, Goodreau, and
  Morris}]{handcock:statnet}
\bibinfo{author}{Handcock, M.~S.}, \bibinfo{author}{Hunter, D.~R.},
  \bibinfo{author}{Butts, C.~T.}, \bibinfo{author}{Goodreau, S.~M.}, and
  \bibinfo{author}{Morris, M.} (\textbf{\bibinfo{year}{2003}}).
  \emph{\bibinfo{title}{statnet: Software tools for the Statistical Modeling of
  Network Data}}, \bibinfo{address}{Seattle, WA},
  \urlprefix\url{http://statnetproject.org}.

\bibitem[{Harris(2014)}]{harris2014introduction}
\bibinfo{author}{Harris, J.} (\textbf{\bibinfo{year}{2014}}).
  \emph{\bibinfo{title}{An Introduction to Exponential Random Graph Modeling}},
  Quantitative Applications in the Social Sciences (\bibinfo{publisher}{SAGE
  Publications}),
  \urlprefix\url{https://books.google.ca/books?id=scQ\_AwAAQBAJ}.

\bibitem[{Heckathorn(1997)}]{Hec97}
\bibinfo{author}{Heckathorn, D.~D.} (\textbf{\bibinfo{year}{1997}}).
  \enquote{\bibinfo{title}{Respondent-driven sampling: a new approach to the
  study of hidden populations.}}, \bibinfo{journal}{Social Problems}
  \textbf{\bibinfo{volume}{44}}, \bibinfo{pages}{174–199}.

\bibitem[{Heckathorn(2002)}]{Hec02}
\bibinfo{author}{Heckathorn, D.~D.} (\textbf{\bibinfo{year}{2002}}).
  \enquote{\bibinfo{title}{Respondent-driven sampling ii: deriving valid
  population estimates from chain-referral samples of hidden populations.}},
  \bibinfo{journal}{Social Problems} \textbf{\bibinfo{volume}{49}},
  \bibinfo{pages}{11--34}.

\bibitem[{Hunter \emph{et~al.}(2008)Hunter, Handcock, Butts, Goodreau, and
  Morris}]{Hunter08}
\bibinfo{author}{Hunter, D.~R.}, \bibinfo{author}{Handcock, M.~S.},
  \bibinfo{author}{Butts, C.~T.}, \bibinfo{author}{Goodreau, S.~M.}, and
  \bibinfo{author}{Morris, M.} (\textbf{\bibinfo{year}{2008}}).
  \enquote{\bibinfo{title}{ergm: A package to fit, simulate and diagnose
  exponential-family models for networks}}, \bibinfo{journal}{Journal of
  Statistical Software} \textbf{\bibinfo{volume}{24}}, \bibinfo{pages}{1--29}.

\bibitem[{Lambert \emph{et~al.}(2019)Lambert, Cox, Messier-Peet, Apelian, and
  Moodie}]{DRSP19}
\bibinfo{author}{Lambert, G.}, \bibinfo{author}{Cox, J.},
  \bibinfo{author}{Messier-Peet, M.}, \bibinfo{author}{Apelian, H.}, and
  \bibinfo{author}{Moodie, E. E.~M.} (\textbf{\bibinfo{year}{2019}}).
  \emph{\bibinfo{title}{Engage Montréal, Portrait de la santé sexuelle des
  hommes de la région métropolitaine de Montréal ayant des relations
  sexuelles avec des hommes, Cycle 2017-2018, Faits saillants}}.

\bibitem[{Lin \emph{et~al.}(2013)Lin, Yeh, and Li}]{lin2013sampling}
\bibinfo{author}{Lin, S.-D.}, \bibinfo{author}{Yeh, M.-Y.}, and
  \bibinfo{author}{Li, C.-T.} (\textbf{\bibinfo{year}{2013}}).
  \enquote{\bibinfo{title}{Sampling and summarization for social networks}}, in
  \emph{\bibinfo{booktitle}{17th Pacific-Asia Conference on Knowledge Discovery
  and Data Mining (PAKDD)(tutorial)}}.

\bibitem[{Newman(2002)}]{Newm02}
\bibinfo{author}{Newman, M.~E.} (\textbf{\bibinfo{year}{2002}}).
  \enquote{\bibinfo{title}{Assortative mixing in networks}},
  \bibinfo{journal}{Physical Review Letters} \textbf{\bibinfo{volume}{89}}.

\bibitem[{Salganik and Heckathorn(2004)}]{Sal04}
\bibinfo{author}{Salganik, M.~J.} and \bibinfo{author}{Heckathorn, D.}
  (\textbf{\bibinfo{year}{2004}}). \enquote{\bibinfo{title}{Sampling and
  estimation in hidden populations using respondent-driven sampling.}},
  \bibinfo{journal}{Sociological Methodology} \textbf{\bibinfo{volume}{34}},
  \bibinfo{pages}{193– 240}.

\bibitem[{Shalizi and Rinaldo(2013)}]{shalizi2013consistency}
\bibinfo{author}{Shalizi, C.~R.} and \bibinfo{author}{Rinaldo, A.}
  (\textbf{\bibinfo{year}{2013}}). \enquote{\bibinfo{title}{Consistency under
  sampling of exponential random graph models}}, \bibinfo{journal}{Annals of
  statistics} \textbf{\bibinfo{volume}{41}}, \bibinfo{pages}{508}.

\bibitem[{Spiller \emph{et~al.}(2018)Spiller, Gile, Handcock, Mar, and
  Wejnert}]{Spil18}
\bibinfo{author}{Spiller, M.~W.}, \bibinfo{author}{Gile, K.~J.},
  \bibinfo{author}{Handcock, M.~S.}, \bibinfo{author}{Mar, C.~M.}, and
  \bibinfo{author}{Wejnert, C.} (\textbf{\bibinfo{year}{2018}}).
  \enquote{\bibinfo{title}{Evaluating variance estimators for respondent-driven
  sampling.}}, \bibinfo{journal}{Journal of Survey Statistics and Methodology}
  \textbf{\bibinfo{volume}{6}}, \bibinfo{pages}{23--45}.

\bibitem[{WHO(2013)}]{WHORDS2013}
\bibinfo{author}{WHO} (\textbf{\bibinfo{year}{2013}}).
  \enquote{\bibinfo{title}{Introduction to hiv/aids and sexually transmitted
  infection surveillance: Module 4: Introduction to respondent-driven
  sampling}}, \bibinfo{type}{Technical Report}.

\end{thebibliography}
\end{document}